\documentclass{article}
\usepackage{spconf,amsmath,graphicx,hyperref}

\renewcommand{\normalsize}{\fontsize{9pt}{11pt}\selectfont}

\makeatletter
\DeclareMathSizes{9}{8}{5.5}{5}
\makeatother

\usepackage[font=footnotesize]{caption}

\usepackage{etoolbox}
\AtBeginEnvironment{thebibliography}{\footnotesize}

\usepackage{cite}

\makeatletter
\def\section{\@startsection{section}{1}{\z@}{13pt plus 2pt minus 1pt}{7pt plus 2pt minus 1pt}{\normalfont\fontsize{10pt}{12pt}\bfseries}}
\def\subsection{\@startsection{subsection}{2}{\z@}{10pt plus 4pt minus 1pt}{5pt plus 3pt minus 1pt}{\normalfont\fontsize{9pt}{11pt}\bfseries}}
\makeatother

\usepackage{etoolbox}
\makeatletter
\patchcmd{\@xsect}{\@afterheading}{\@afterheading\@afterindentfalse}{}{}
\makeatother

\usepackage{tipa}
\usepackage{float}
\usepackage{amsmath}
\usepackage{amssymb}
\usepackage{multirow}
\usepackage{booktabs}
\usepackage{setspace}
\usepackage{caption}
\title{MECT: Mixture of Experts with CNN-Transformer Network for Speaker verification}
\name{Yu Zheng, Jinghan Peng, ChangHao Zhang, Jian Liu$^\dagger$\thanks{$^\dagger$Corresponding author: rex.lj@antgroup.com\\ 
\url{https://github.com/ant-research/AntSpeaker}}, Weiqiang Wang}
\address{Machine Intelligence, Ant Group, Shanghai, China}
\begin{document}
%\ninept
%
\maketitle
\begin{abstract}
In this paper, we propose MECT, a speaker verification model that integrates the Mixture-of-Experts (MoE) mechanism into a CNN-Transformer backbone with optimized block structure and stacking scheme. Specifically, we investigated four MoE variants that span utterance-level and frame-level granularity with dense and sparse routing strategies. The MoE mechanism proves to be effective over the baseline without MoE with only a small increase in parameters. We further scale MECT to a series of model sizes, all maintaining compact parameters and low computational complexity. In particular, MECT-B2 achieves state-of-the-art performance on VoxCeleb1 and delivers strong results on CN-Celeb, demonstrating its effectiveness across diverse datasets. In addition, we establish a streaming inference paradigm through causal retraining, which maintains strong performance at a chunk size of 100ms.
\end{abstract}
\section{Introduction}
\label{sec:intro}
Speaker verification (SV) aims to determine whether two speech segments are from the same speaker. In recent years, deep neural network architectures for SV have evolved rapidly. Representative models include 1D convolution TDNN-based frameworks such as ECAPA-TDNN~\cite{ecapatdnn,pcf} and Next-TDNN~\cite{nexttdnn}, CNN-based architectures such as ResNet~\cite{resnet,resnet-but} and its variants~\cite{resnext,res2net} including DF-ResNet and Gemini DF-ResNet~\cite{df-resnet,gemini}, and CNN--TDNN hybrid structures such as CAM++~\cite{cam} and ECAPA2~\cite{ecapa2}. 
Transformer-based networks have also been investigated~\cite{mfa-conformer,transformer-speaker,cnn-transformer,redimnet}. 
ReDimNet~\cite{redimnet} introduces a reshape-dimension strategy to effectively integrate CNN, TDNN, and Transformer modules, while ReDimNet2~\cite{redimnet2} further improves performance through time-pooled dimension reshaping. Self-supervised pre-trained models (PTMs) have also been applied to speaker verification, with w2v-BERT 2.0-based SV~\cite{w2v-bert} achieving particularly strong performance through fine-tuning.
% 增加一句铺垫，衔接之后的MoE
% Despite these advances, the feed-forward block in existing SV architectures typically relies on a single transformation path, which limits the diversity of learned representations and constrains further performance gains.

Mixture-of-Experts (MoE)~\cite{moe} has been widely adopted in large-scale Transformer-based systems, improving scalability through dynamic expert routing. In speaker verification, MoE has only been applied to fine-tuning PTMs~\cite{moe-ptm,ptm-moe}, where it adaptively combines representations from different layers of the pretrained model. 
However, MoE has not yet been explored in fully supervised SV.
% To the best of our knowledge, MoE has not yet been explored in fully supervised SV.

We propose MECT, a fully supervised speaker verification model that introduces the MoE mechanism. Four MoE structures are designed with different expert granularities and routing strategies: utterance-level Dense MoE, utterance-level Sparse MoE, frame-level Dense MoE, and frame-level Sparse MoE, each replacing the first linear projection in the Transformer feed-forward block. 
MECT adopts a CNN-Transformer architecture based on the Reshape Dimension strategy of ReDimNet, with a redesigned stacking scheme and network structure for better performance. 
The main contributions of this paper are summarized as follows:

1) We propose MECT, the first MoE-based fully supervised speaker verification model, achieving state-of-the-art performance on Vox1-O, Vox1-E, and Vox1-H with minDCF of 0.012, 0.026, and 0.048, while maintaining low model complexity. 2) We investigate four MoE structures and demonstrate performance gains over the baseline across datasets, with frame-level dense MoE excelling on VoxCeleb and sparse Top-K MoE on CN-Celeb. 3) We establish a streaming inference paradigm for SV through causal retraining, achieving competitive performance at a chunk size of 100ms.

\begin{figure*}[t]
 \centering
 \includegraphics[width=0.94\textwidth]{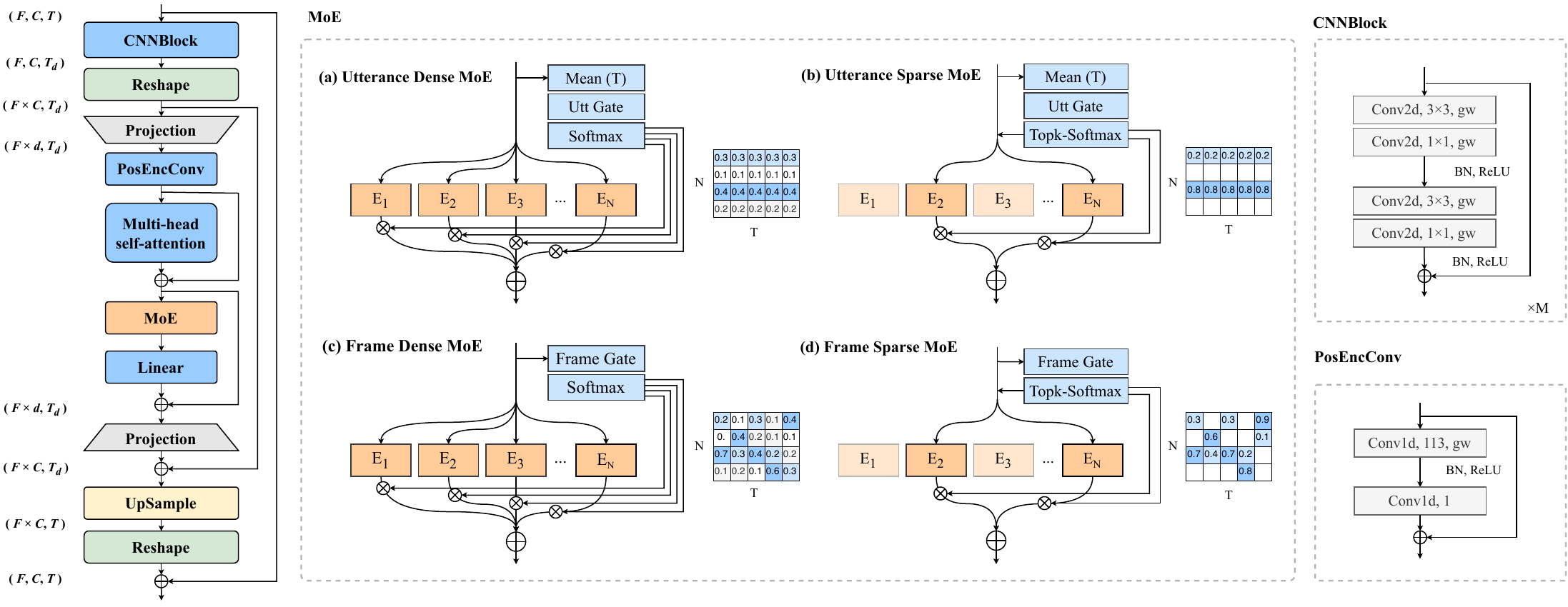}
 \caption{\textbf{The architecture of MECT Block}. Left shows the overall block, middle the four MoE variants, and right the CNN and PosEncConv modules.}
 \label{mectblock}
 \vspace{-0.5cm}
\end{figure*}

\section{METHODS}
\label{sec:format}

This section describes the architecture of MECT. 
We first introduce the basic building block, then detail the MoE layer as the core component and a key contribution of this work. 
% Next, we present how MECT blocks are stacked to construct a series of models at different scales. 
Next, we present the stacking scheme for constructing MECT models at different scales to accommodate diverse deployment requirements. % 增加一句说明为什么要不同scale的模型型号
Finally, we describe the streaming speaker embedding extraction scheme.

\subsection{Basic Block of MECT}
\label{ssec:subhead}
% As illustrated in Figure\ref{mectblock}, the block of MECT consists of a CNN and a Transformer equipped with MoE. 
As illustrated in Figure\ref{mectblock}, a MECT block consists of a CNN and a Transformer equipped with MoE. 
The CNN and Transformer are connected via the reshape and time-pooled dimension-reshaping methods adopted from ReDimNet and ReDimNet2\cite{redimnet,redimnet2}. 
Beyond MoE, our block design also differs from ReDimNet in several aspects that contribute to further performance gains. % 预告与 redimnet 的区别

\textbf{Residual connection}. We adopt ResNet blocks for the CNN, but place the shortcut addition after BatchNorm and ReLU rather than before, as shown in the CNNBlock module of Figure\ref{mectblock}. 
This design consistently yields better performance in our experiments.

\textbf{Position encoding.} We employ a two-layer 1D convolution as the position encoding, as illustrated in the PosEncConv of Figure\ref{mectblock}.
% Compared with sinusoidal encodings, convolution-based position encoding better captures local temporal patterns, which is beneficial for speaker-discriminative features. % 增加一句说明为什么用 1dconv 位置编码

\textbf{Downsampling and channel expansion.} 
Both temporal downsampling and channel expansion are handled by convolutions within the ResNet block, rather than by a separate convolution layer before it as in ReDimNet. 
% This simplifies the module structure and yields more stable training.
This design simplifies the module structure and leads to more stable training.

\textbf{Block connection.} 
% We connect blocks via a simple shortcut rather than the dense connections used in ReDimNet, which did not yield any benefit in our experiments.
We connect blocks via a simple shortcut rather than the dense connections used in ReDimNet. In our experiments, dense connections did not improve performance while increasing computational overhead.

\subsection{Mixture-of-Experts layer}
\label{ssec:subhead}
We explore four MoE variants that differ in aggregation granularity (utterance-level vs.\ frame-level) and routing strategy (Dense vs.\ Sparse), as shown in Figure~\ref{mectblock}.

Let $\mathbf{x} \in \mathbb{R}^{T \times d}$ denote the input feature sequence to the MoE layer, where $T$ and $d$ are the temporal length and feature dimension, respectively. Let $\mathbf{x}_t \in \mathbb{R}^{d}$ denotes a single frame.

\subsubsection{Frame-level MoE}
\label{sssec:subsubhead}
In the Dense MoE, the gating network is a linear layer parameterized by $\mathbf{W}_g \in \mathbb{R}^{d \times N}$, where $N$ denotes the number of experts. 
Its output $\mathbf{G}_t$ serves as the routing weights for each expert at frame $t$. 
Each expert is a linear layer parameterized by $\mathbf{W}_e^{i}$, and $f(\cdot)$ denotes the activation function. 
% The computation is formulated as follows:
The output of the MoE layer, $\mathbf{y}_t$, is computed as
% \vspace{-8pt}
\begin{equation}
\mathbf{G}_t = \text{Softmax}(\mathbf{x}_t \mathbf{W}_g), \quad
\mathbf{y}_t = \sum_{i=1}^{N} \mathbf{G}_t^{i} \cdot f(\mathbf{x}_t \mathbf{W}_e^{i}).
\label{eq:dense_output}
\end{equation}
In the Sparse MoE, only the Top-$K$ experts are activated for routing. $\mathcal{S}_t$ and $\mathbf{h}_t$ denote the indices and the corresponding logits of the selected experts, respectively. It is formulated as follows:
% \vspace{-4pt}
\begin{equation}
\mathcal{S}_t, \mathbf{h}_t = \text{TopK}(\mathbf{x}_t \mathbf{W}_g, K), \quad
\label{eq:sparse_gate}
\end{equation}
\begin{equation}
\mathbf{G}_t = \text{Softmax}(\mathbf{h}_t), \quad
\mathbf{y}_t = \sum_{i \in \mathcal{S}_t} \mathbf{G}_t^{i} \cdot f(\mathbf{x}_t \mathbf{W}_e^{i}).
\label{eq:sparse_output}
\end{equation}

\subsubsection{Utterance-level MoE}
\label{sssec:subsubhead}
% In the utterance-level variants, the gating decision is made based on the mean-pooled utterance representation $\bar{\mathbf{x}}$, producing a single routing weight shared across all frames.
In the utterance-level variants, the gating decision is made based on the mean-pooled utterance representation $\bar{\mathbf{x}} = \frac{1}{T}\sum_{t=1}^{T} \mathbf{x}_t$, producing a single set of routing weights shared across all frames.
The utterance-level Dense MoE is formulated as follows:
% \begin{equation}
% \bar{\mathbf{x}} = \frac{1}{T}\sum_{t=1}^{T} \mathbf{x}_t
% \label{eq:utt_dense_pool}
% \end{equation}
% \vspace{-4pt}
\begin{equation}
\mathbf{G} = \text{Softmax}(\bar{\mathbf{x}} \mathbf{W}_g), \quad
\mathbf{y}_t = \sum_{i=1}^{N} \mathbf{G}^{i} \cdot f(\mathbf{x}_t \mathbf{W}_e^{i}).
\label{eq:utt_dense_output}
\end{equation}
And the utterance-level Sparse MoE is formulated as follow:
\begin{equation}
\mathcal{S}, \mathbf{g} = \text{TopK}(\bar{\mathbf{x}} \mathbf{W}_g, K), %\quad
% \mathbf{G} = \text{Softmax}(\mathbf{g})
\label{eq:utt_sparse_gate}
\end{equation}
\begin{equation}
\mathbf{G} = \text{Softmax}(\mathbf{g}), \quad
\mathbf{y}_t = \sum_{i \in \mathcal{S}} \mathbf{G}^{i} \cdot f(\mathbf{x}_t \mathbf{W}_e^{i}).
\label{eq:utt_sparse_output}
\end{equation}

\subsection{MECT architecture}
\label{ssec:subhead}

\begin{table}[!t]
\centering
\footnotesize
% \small
% \setlength{\tabcolsep}{6pt}
\caption{\textbf{Architecture of MECT.}}
\label{mctnet-structure}
\renewcommand{\arraystretch}{0.95} % 控制行高（行间距）
\begin{tabular}{ccc}
\toprule
\textbf{Layer name} & \textbf{Structure}         & \textbf{Output} \\ \midrule
Input               & -                          & (1, 80, T )    \\
Head              & Conv2D, stride 1       & (C, 80, T)      \\ \midrule
Stage1       & Block × 2, stride 1 & (C, 80, T)      \\
Stage2       & Block × 4, stride (2,1),(1,2) × 3 & (2 × C, 40, T)    \\
Stage3       & Block × 5, stride (2,1),(1,2) × 4 & (4 × C, 20, T)    \\
Stage4       & Block × 3, stride (2,1),(1,2) × 2 & (8 × C, 10, T)    \\ \midrule
Fusion        & Weight Dense         & (80 × C, T)       \\ \midrule
Pooling             & ASTP                       & 80 × C            \\
Projection           & Linear                     & 192             \\ \midrule
Loss                & SphereFace2                & Num. Speakers               \\ 
\bottomrule
\end{tabular}
\end{table}

Table~\ref{mctnet-structure} details the MECT architecture. Each stage stacks multiple MECT blocks, where the first block applies frequency 
downsampling, and the remaining blocks apply temporal downsampling, both via convolutional stride. Outputs of all stages are fused via a weighted dense layer before pooling.

We provide four MECT model sizes controlled by two parameters: the input channel size $C$ and the number of ResBlocks per block $M$, with 
a fixed Transformer hidden dimension $d=64$. A1 and B1 use $M=2$, while A2 and B2 use $M=3$; A1 and A2 use $C=16$, while B1 and B2 use $C=32$.

% We use 80-dimensional mean-normalized log Mel filter-bank features extracted with a 25\,ms window and 10\,ms frame shift over 20--7600\,Hz. 
We use 80-dimensional mean-normalized log Mel filter-bank features extracted with a 25\,ms window and 10\,ms frame shift over 20--7600\,Hz at a sampling rate of 16\,kHz. 
% Attentive Statistics Pooling (ASTP)~\cite{astp} and SphereFace2 loss~\cite{sphereface2} are employed.
 Attentive Statistics Pooling (ASTP)~\cite{astp} is employed for utterance-level aggregation, and SphereFace2 loss~\cite{sphereface2} is used for model training.
 
\subsection{Streaming speaker embedding extract}
\label{ssec:subhead}

We compare three streaming strategies, denoted M1, M2, and M3.
M1 (embedding averaging) simply averages the embedding of the current chunk with those of all prior chunks. 
M2 (feature concatenation) concatenates all accumulated features before pooling, then passes them through the subsequent network to produce the embedding. 
M3 (causal retraining) requires retraining the model under a causal architecture, as described below.

We first convert the network to a causal architecture. 
For convolution layers, padding is applied only on the left side of the temporal 
dimension, and for multi-head attention (MHA), a causal mask restricts each frame to attend only to preceding frames. 
The model is retrained accordingly.

% During streaming inference, as illustrated in Figure~\ref{streaming}, 
% Conv2d layers cache 2 frames (or 6 when stride of temporal dimensionis is 2) of the feature 
% input from the previous step for receptive field alignment, while 
% Conv1d layers cache 112 frames to cover the accumulated stride. These 
% cached frames are concatenated with the current input feature along the 
% temporal dimension. For MHA, the KV states of all previous steps are 
% concatenated with the current step before attention. The MECT output of 
% prior steps is also cached and pooled together with the current output. 
% Since only the KV cache and MECT output pooling rely on the full 
% history, and both operations are computationally lightweight, the 
% additional cost remains modest. In practice, the KV cache and MECT output cache can be capped to bound 
% memory usage. Our experiments do not apply this cap, as the focus 
% is on validating the streaming algorithm.

\begin{figure}[t]
  \centering
  \includegraphics[width=0.42\textwidth,height=0.25\textwidth]{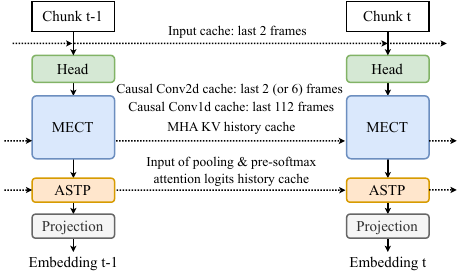}
  \caption{\textbf{Streaming inference pipeline of MECT}. Illustrating the causal processing and multi-level caching across consecutive chunks.}
  \label{streaming}
\end{figure}

During streaming inference (Figure~\ref{streaming}), no mean normalization is applied to the input fbank features. Conv2d layers cache 2 frames (or 6 for temporal stride 2) for receptive field alignment, while Conv1d layers cache 112 frames to cover the accumulated stride, both concatenated with the current input along the temporal dimension. For MHA, the KV states of all previous steps are concatenated with the current step before attention. For ASTP, the input frame features and intermediate attention logits (feature before softmax) from prior chunks are cached and concatenated with the current chunk to compute global mean and standard deviation. Since only the KV cache and ASTP rely on full history and both are lightweight, the additional cost remains modest. In practice, both caches can be capped to bound memory usage. Our experiments do not apply this cap to focus on validating the streaming algorithm.

\section{Experiment}
\label{sec:pagestyle}

\subsection{Datasets}
\label{ssec:subhead}

We evaluated MECT on multiple datasets. 
For VoxCeleb model training, we used the VoxCeleb2~\cite{vox2} development set, either alone or combined with VoxBlink2~\cite{voxblink2}. 
SV performance was evaluated with VoxCeleb1-Cleaned protocols: Vox1-O, Vox1-E, Vox1-H and Vox21-val. 
Additionally, the development sets of CN-Celeb1 and CN-Celeb2~\cite{cnceleb} were used for training, with evaluation on CN-Celeb Test. 
These experiments span different languages, scales, and data sources, validating the effectiveness of the proposed method.

\begin{table}[t]
\centering
\footnotesize
\renewcommand{\arraystretch}{0.85}
\setlength{\tabcolsep}{2.8pt}
\caption{\textbf{Evaluation results on different MoE types.}}
\label{moe_type}
\begin{tabular}{lccccccc}
\toprule
\multirow{2}{*}{\textbf{Type}} & \multirow{2}{*}{\textbf{No.E}} & \multirow{2}{*}{\textbf{Vox1-O}} & \multirow{2}{*}{\textbf{Vox1-E}} & \multirow{2}{*}{\textbf{Vox1-H}} & \multirow{2}{*}{\textbf{Vox21-val}} & \multicolumn{2}{c}{\textbf{Avg}} \\ %\cline{7-8}
 & & & & & & \textbf{EER} & \textbf{DCF} \\ \midrule
No MoE & -- & 0.33 & 0.48 & 0.85 & 1.68 & 0.84 & 0.064 \\ \midrule
F-Dense & 2 & 0.34 & 0.49 & 0.89 & 1.71 & 0.86 & 0.066 \\
F-Dense & 3 & \textbf{0.27} & 0.49 & 0.89 & 1.65 & 0.83 & 0.061 \\
F-Dense & 4 & \textbf{0.27} & \textbf{0.46} & \textbf{0.85} & 1.60 & \textbf{0.80} & \textbf{0.059} \\
F-Dense & 5 & 0.35 & 0.51 & 0.88 & 1.61 & 0.84 & 0.061 \\
% U-Dense & 2 & 0.35 & 0.50 & 0.91 & 1.68 & 0.86 & 0.064 \\
U-Dense & 3 & 0.30 & 0.47 & 0.87 & 1.62 & 0.82 & 0.062 \\
U-Dense & 4 & 0.36 & 0.48 & 0.89 & 1.60 & 0.83 & 0.067 \\ \midrule
% U-Dense & 5 & 0.34 & 0.48 & 0.88 & 1.67 & 0.84 & 0.064 \\ \midrule
% F-Top1 & 2 & 0.32 & 0.47 & 0.87 & 1.60 & 0.82 & 0.063 \\
F-Top1 & 4 & 0.36 & 0.48 & 0.88 & 1.61 & 0.83 & 0.064 \\
% U-Top1 & 2 & 0.32 & 0.50 & 0.89 & 1.62 & 0.83 & 0.064 \\
U-Top1 & 4 & 0.37 & 0.48 & 0.87 & 1.61 & 0.83 & 0.061 \\ \midrule
% F-Top2 & 6 & 0.29 & \textbf{0.45} & \textbf{0.85} & 1.63 & 0.81 & 0.061 \\
% F-Top2 & 8 & 0.30 & 0.47 & 0.86 & 1.67 & 0.83 & 0.063 \\
F-Top4 & 6 & 0.33 & 0.49 & 0.89 & 1.57 & 0.82 & 0.066 \\
F-Top4 & 8 & 0.30 & 0.48 & 0.87 & 1.55 & 0.80 & 0.063 \\
F-Top4 & 16 & 0.31 & 0.47 & 0.86 & 1.57 & 0.80 & 0.061 \\
F-Top4 & 32 & 0.33 & 0.48 & 0.88 & \textbf{1.51} & 0.80 & 0.063 \\
\bottomrule
\end{tabular}
\end{table}

\begin{table*}[t]
\centering
\footnotesize
\caption{\textbf{Evaluation results on VoxCeleb1-Cleaned.} * indicates results obtained using the VoxCeleb2 and VoxBlink2 datasets for training. GMACs were measured on 2-s segments counting only the backbone, with ReDimNet2 using the official open-source model.}
\label{results}
\renewcommand{\arraystretch}{0.85}
\begin{tabular}{lcccccccccc}
\toprule
\multirow{2}{*}{\textbf{Model}} & \multirow{2}{*}{\textbf{Params}} & \multirow{2}{*}{\textbf{GMACs}} & \multirow{2}{*}{\textbf{LMF}} & \multirow{2}{*}{\textbf{AS-Norm}} & \multicolumn{2}{c}{\textbf{Vox1-O}} & \multicolumn{2}{c}{\textbf{Vox1-E}} & \multicolumn{2}{c}{\textbf{Vox1-H}} \\ %\cline{6-11}
 & & & & & \textbf{EER(\%)} & \textbf{minDCF} & \textbf{EER(\%)} & \textbf{minDCF} & \textbf{EER(\%)} & \textbf{minDCF} \\ \midrule
ECAPA (C=512)~\cite{ecapatdnn,df-resnet} & 6.2M & 1.04 & $\times$ & \checkmark & 0.94 & 0.092 & 1.21 & 0.129 & 2.20 & 0.205 \\
CAM++~\cite{cam,wespeaker} & 7.2M & 1.15 & \checkmark & $\times$ & 0.71 & 0.109 & 0.85 & 0.095 & 1.66 & 0.165 \\
ReDimNet2-B3~\cite{redimnet2} & 2.58M & 2.58 & \checkmark & $\times$ & \textbf{0.42} & 0.038 & 0.66 & 0.068 & 1.22 & 0.120 \\
\textbf{MECT-A1} & 3.78M & 3.02 & \checkmark & $\times$ & 0.44 & \textbf{0.038} & \textbf{0.60} & \textbf{0.060} & \textbf{1.05} & \textbf{0.108} \\ \midrule
Gemini DF-ResNet114~\cite{gemini} & 6.5M & 5.0 & $\times$ & \checkmark & 0.69 & 0.067 & 0.86 & 0.097 & 1.49 & 0.144 \\
ReDimNet2-B4~\cite{redimnet2} & 4.52M & 4.48 & \checkmark & $\times$ & 0.37 & 0.040 & 0.58 & 0.060 & 1.07 & 0.102 \\
\textbf{MECT-A2} & 4.12M & 4.11 & \checkmark & $\times$ & \textbf{0.37} & \textbf{0.032} & \textbf{0.54} & \textbf{0.056} & \textbf{0.97} & \textbf{0.096} \\ \midrule
ResNet293~\cite{wespeaker} & 28.6M & 28.10 & \checkmark & $\times$ & 0.53 & 0.057 & 0.71 & 0.072 & 1.30 & 0.127 \\
ECAPA2~\cite{ecapa2} & 27.1M & 187.0 & \checkmark & $\times$ & 0.44 & 0.041 & 0.62 & 0.066 & 1.15 & 0.114 \\
WavLM Large~\cite{kim25i_interspeech} & 319.3M & 28.5 & \checkmark & \checkmark & 0.37 & 0.059 & 0.50 & 0.055 & 1.01 & 0.099 \\
\textbf{MECT-B1} & 8.26M & 10.35 & \checkmark & $\times$ & 0.37 & 0.029 & 0.50 & 0.051 & 0.94 & 0.092 \\
ReDimNet2-B6~\cite{redimnet2} & 10.45M & 13.11 & \checkmark & $\times$ & 0.29 & 0.027 & 0.52 & 0.050 & 0.99 & 0.104 \\
\textbf{MECT-B2} & 9.57M & 14.43 & \checkmark & $\times$ & \textbf{0.27} & \textbf{0.024} & \textbf{0.46} & \textbf{0.048} & \textbf{0.85} & \textbf{0.082} \\ \midrule
W2V-BERT 2.0*~\cite{w2v-bert} & \multirow{2}{*}{587M} & \multirow{2}{*}{57.90} & \checkmark & $\times$ & 0.14 & 0.020 & 0.31 & 0.032 & 0.73 & 0.071 \\
\multicolumn{1}{r}{+AS-Norm \& QMF} & & & \checkmark & \checkmark & \textbf{0.12} & 0.025 & \textbf{0.27} & 0.028 & 0.55 & 0.051 \\
\textbf{MECT-B2}* & \multirow{2}{*}{9.57M} & \multirow{2}{*}{14.43} & \checkmark & $\times$ & 0.23 & 0.013 & 0.29 & 0.028 & 0.54 & 0.052 \\
\multicolumn{1}{r}{+AS-Norm \& QMF} & & & \checkmark & \checkmark & 0.22 & \textbf{0.012} & 0.28 & \textbf{0.026} & \textbf{0.52} & \textbf{0.048} \\
\bottomrule
\end{tabular}
\vspace{-0.1cm}
\end{table*}

\subsection{Training}
\label{ssec:subhead}

Experiments were conducted under three data settings. All models were trained in two stages: full training followed by large-margin fine-tuning (LMF)~\cite{lmf}. All experiments used the same hardware and PyTorch environment, and random seeds were fixed across ablation studies to isolate the effect of each component.

\textbf{VoxCeleb2.} VoxCeleb2 was augmented with two-fold speed perturbation (0.9 and 1.1)~\cite{augment}, yielding 17,982 speakers. Two-second segments were randomly extracted with noise and reverberation augmentation via MUSAN~\cite{musan} and RIR~\cite{rir} following~\cite{wespeaker}. SGD was used with momentum 0.9, weight decay 2e-5, and batch size 512. The learning rate warmed up to 0.6 over 4 epochs, then decayed to 4e-4 at epoch 120. The SphereFace2 margin was set to 0.0 for 20 epochs, raised to 0.2 over the next 20, and held constant. In the LMF stage, speed perturbation was removed (5,994 speakers), segment length increased to 6\,s, the margin was fixed at 0.3, and the learning rate was reduced to 2e-4 for 7 epochs.

\textbf{CN-Celeb.} The same training recipe was adopted with a maximum learning rate of 1.6. The model from epoch 50 of the first stage was used for LMF, with a learning rate of 8e-4.

\textbf{VoxCeleb2 \& VoxBlink2.} Speed perturbation was not applied. The first-stage training followed the same settings as the VoxCeleb2 experiment, while the second stage used VoxCeleb2 alone for LMF.

\subsection{Evaluation}
\label{ssec:subhead}
The performance was evaluated using Equal Error Rate (EER) and minimum Decision Cost Function (minDCF) with $C_{\mathrm{FA}}=C_{\mathrm{Miss}}=1$, $P_{\mathrm{target}}=0.01$, except on Vox21-val $P_{\mathrm{target}}=0.05$. 
% AS-Norm~\cite{asnorm} and QMF~\cite{lmf} were applied for certain systems using VoxCeleb2 data as backend.
AS-Norm~\cite{asnorm} and QMF~\cite{lmf} were applied for selected systems using VoxCeleb2 development set as backend.

\subsection{Streaming inference test}
\label{ssec:subhead}
During streaming evaluation, the input audio was processed chunk by chunk, and the final embedding was obtained after the last chunk. For M1 and M2, adjacent chunks overlap by half the chunk size for better performance.

\section{Results}
\label{sec:typestyle}
% \subsection{MoE Ablation Study}
Table~\ref{moe_type} reports representative results from our full 
ablation study of different MoE types and expert counts, all trained on 
VoxCeleb2 and evaluated on the VoxCeleb1 test sets. ``No MoE'' denotes 
the baseline without MoE modules. Among all configurations, Frame-level 
Dense MoE with 4 experts achieves the best results, outperforming the 
baseline across all test sets with relative gains of 4.7\% and 7.8\% on 
average EER and minDCF. Notably, this improvement comes at a negligible 
parameter cost: the baseline contains 9.40M parameters versus 9.57M for 
the best MoE configuration.

\begin{figure}[t]
  \centering
  \includegraphics[width=0.46\textwidth]{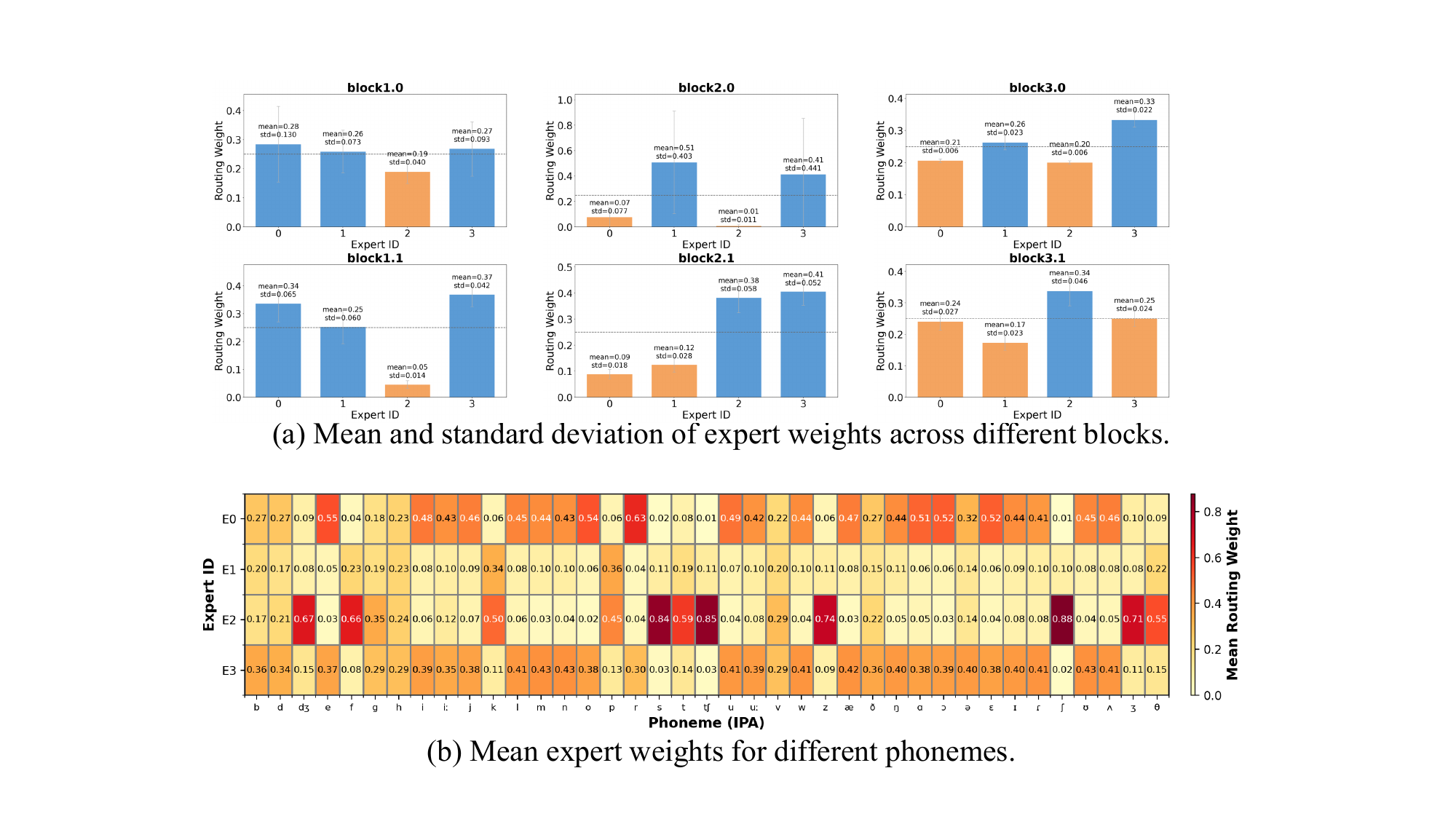}
  % \caption{\textbf{The architecture of MCTBlock}. The left panel illustrates.}
  % \caption{Statistics of expert routing weights across network blocks and phonemes.}
  \caption{\textbf{Statistics of expert weights across network blocks and phonemes.}}
  \label{moe_weight}
% \vspace{-0.1cm}
\end{figure}

% Table~\ref{results} compares MECT with publicly available models on the 
% VoxCeleb1 protocols, grouped by parameter count and MACs. In the second 
% group, MECT-A2 achieves the best results with the fewest parameters and 
% MACs, outperforming ReDimNet2-B4 by 6.0\% and 9.0\% on average EER and 
% minDCF. In the third group, MECT-B2 surpasses ReDimNet2-B6 by 11.7\% 
% and 15.0\% on average EER and minDCF, despite having fewer parameters 
% and only slightly higher MACs, while the even smaller MECT-B1 already 
% edges past ReDimNet2-B6. The fourth group adds VoxBlink2 training, 
% where MECT-B2 outperforms w2v-bert2.0 with far fewer parameters and 
% MACs, achieving better EER and minDCF on Vox1-E and Vox1-H, and a 
% notably lower minDCF of 0.013 on Vox1-O. Overall, MECT-B2 yields 
% relative gains of 10.3\% and 24.4\% on average EER and minDCF.
% Table~\ref{cnceleb-results} reports results on CN-Celeb. MECT-B2 with frame-level Dense MoE 
% (4 experts) continues to perform well, achieving a 5.2\% relative EER 
% improvement over ReDimNet2-B6 trained under the same settings.

Table~\ref{results} compares MECT with publicly available models on the VoxCeleb1 protocols, where MECT uses frame-level Dense MoE with 4 experts. 
In group 2, MECT-A2 outperforms ReDimNet2-B4 across all three test sets with the fewest parameters and MACs, achieving particularly notable gains on Vox1-H (9.4\% and 5.9\% relative improvements on EER and minDCF). 
In group 3, MECT-B2 surpasses ReDimNet2-B6 with fewer parameters and comparable MACs, delivering consistent improvements on all sets and a 14.1\% relative gain on Vox1-H minDCF, while the smaller MECT-B1 already edges past ReDimNet2-B6. With VoxBlink2 training, MECT-B2 outperforms w2v-bert2.0 using far fewer resources, achieving substantial improvements on Vox1-O, Vox1-E, and Vox1-H with relative gains of 10.3\% and 24.4\% on average EER and minDCF. 
After AS-Norm and QMF, MECT-B2 further improves and reaches state-of-the-art performance, with minDCFs of 0.012, 0.026, and 0.048. 
Table~\ref{cnceleb-results} reports results on CN-Celeb with models trained only on CN-Celeb. MECT-B2 uses frame-level sparse Top-4 MoE with 8 experts and achieves relative improvements of 6.7\% on EER and 7.5\% on minDCF over ReDimNet2-B6 trained under the same settings.

% Figure~\ref{moe_weight} visualizes the MoE routing behavior in MECT-B2 
% on VoxCeleb1. Using the Allosaurus~\cite{allosaurus} phoneme recognizer, we correlate 
% routing weights with phoneme categories in the first block. As shown in 
% Figure~\ref{moe_weight}(b), E2 consistently assigns high weights 
% (0.65--0.88) to fricatives and affricates~\cite{pickett2000acoustics} (e.g., /s, z, \textesh, f, 
% \texttheta, t\textesh, d\textyogh/), whose high-frequency noise 
% distinguishes them acoustically from other phonemes, while vowels 
% (e.g., /e, o, u, \ae/) are routed mainly to E0 with moderate weights 
% (0.40--0.55). This phoneme-class specialization emerges without any 
% phoneme supervision, suggesting that the MoE mechanism naturally 
% captures the underlying acoustic structure of speech.

Figure\ref{moe_weight} shows the MoE routing behavior in MECT-B2 on VoxCeleb1. Using the Allosaurus\cite{allosaurus} phoneme recognizer, we correlate routing weights with phoneme categories in the first block. As shown in Figure\ref{moe_weight}(b), E2 assigns high weights (0.65--0.88) to fricatives and affricates\cite{pickett2000acoustics} (e.g., /s, z, \textesh, f, \texttheta, t\textesh, d\textyogh/), whose high-frequency noise distinguishes them from other phonemes, while vowels (e.g., /e, o, u, \ae/) are routed mainly to E0 with moderate weights (0.40--0.55). This specialization emerges without phoneme supervision, suggesting the MoE mechanism naturally captures the acoustic structure of speech.

% As shown in Table~\ref{streaming_results}, at a chunk size of 0.1\,s, M3 achieves EERs of 0.38\%, 0.60\%, and 1.06\% on Vox1-O, Vox1-E, and Vox1-H, dramatically outperforming M1 and M2, and remaining close to its offline baseline. 
% This demonstrates the effectiveness of causal retraining for short-chunk streaming.

As shown in Table~\ref{streaming_results}, at a chunk size of 1.0s, M2 achieves the best streaming performance. However, at 0.1s, M1 and M2 degrade severely to 12.34\% and 7.50\% on Vox1-O, while M3 remains at 0.38\%, 0.60\%, and 1.06\% on Vox1-O, Vox1-E, and Vox1-H, closely matching its offline baseline. This demonstrates the effectiveness of causal retraining for short-chunk streaming.

\begin{table}[t]
\centering
\footnotesize
\renewcommand{\arraystretch}{0.85}
\setlength{\tabcolsep}{2.4pt}
\caption{\textbf{Evaluation results on CN-Celeb.}}
\label{cnceleb-results}
\begin{tabular}{lcccccc}
\toprule
\multirow{2}{*}{\textbf{Model}} & \multirow{2}{*}{\textbf{Params}} & \multirow{2}{*}{\textbf{GMACs}} & \multirow{2}{*}{\textbf{LMF}} & \multirow{2}{*}{\textbf{AS-Norm}} & \multicolumn{2}{c}{\textbf{CN-Celeb Test}} \\ %\cline{6-7}
 & & & & & \textbf{EER(\%)} & \textbf{DCF} \\ \midrule
CAM++~\cite{3d-speaker} & 7.2M & 1.15 & \checkmark & $\times$ & 6.30 & 0.370 \\
ERes2NetV2~\cite{3d-speaker} & 17.8M & 12.60 & \checkmark & $\times$ & 6.04 & 0.362 \\
ResNet221~\cite{wespeaker} & 23.9M & 21.29 & \checkmark & \checkmark & 5.66 & 0.330 \\
ReDimNet2-B6 & 12.3M & 13.05 & \checkmark & $\times$ & 5.22 & 0.321 \\
\textbf{MECT-B2} & 9.81M & 14.46 & \checkmark & $\times$ & \textbf{4.87} & \textbf{0.297} \\
 \multicolumn{1}{r}{w/o MoE} & 9.57M & 14.43 & \checkmark & $\times$ & 5.01 & 0.303 \\
% \multicolumn{1}{r}{+w/o MoE} & 9.40M & 14.40 & \checkmark & $\times$ & & \\
\bottomrule
\end{tabular}
\end{table}

\begin{table}[t]
\centering
\footnotesize
\renewcommand{\arraystretch}{0.85}
\setlength{\tabcolsep}{3pt}
\caption{\textbf{Streaming inference results on VoxCeleb1.} MECT-B2$^\dagger$ denotes the causally retrained model.}
\label{streaming_results}
\begin{tabular}{llcccc}
\toprule
\textbf{Model} & \textbf{Method} & \textbf{Chunk} & \textbf{Vox1-O} & \textbf{Vox1-E} & \textbf{Vox1-H} \\ \midrule
\multirow{5}{*}{MECT-B2} & Offline & -- & 0.27 & 0.46 & 0.85 \\ \cmidrule{2-6}
 & \multirow{2}{*}{Stream M1} & 0.1s & 12.34 & 12.36 & 17.40 \\
 & & 0.2s & 4.50 & 4.45 & 6.87 \\
 & & 1.0s & 0.47 & 0.67 & 1.15 \\ \cmidrule{2-6}
 & \multirow{2}{*}{Stream M2} & 0.1s & 7.50 & 7.55 & 10.91 \\
 & & 0.2s & 2.04 & 2.14 & 3.50 \\
 & & 1.0s & 0.39 & 0.58 & 1.03 \\ \midrule
\multirow{2}{*}{MECT-B2$^\dagger$} & Offline & -- & 0.38 & 0.59 & 1.05 \\ \cmidrule{2-6}
 & Stream M3 & 0.1s & 0.38 & 0.60 & 1.06 \\
\bottomrule
\end{tabular}
\end{table}

%\vfill
%\pagebreak

\section{Conclusions}
\label{sec:copyright}

We proposed MECT, a MoE-based speaker verification model. Four MoE structures were investigated, and performance gains over the baseline were observed across datasets. MECT achieves state-of-the-art performance on VoxCeleb1. 
A causal retraining strategy enables streaming inference at 0.1\,s chunks with near-offline accuracy, substantially outperforming non-causal streaming methods. 
Visualization of expert routing indicates that MoE learns phoneme-level specialization to some extent, without explicit supervision.

\vfill\pagebreak

% \section{REFERENCES}
% \label{sec:refs}

% List and number all bibliographical references at the end of the
% paper. The references can be numbered in alphabetic order or in
% order of appearance in the document. When referring to them in
% the text, type the corresponding reference number in square
% brackets as shown at the end of this sentence \cite{C2}. An
% additional final page (the fifth page, in most cases) is
% allowed, but must contain only references to the prior
% literature.

\bibliographystyle{IEEEbib}
\bibliography{strings,refs}

@inproceedings{resnet,
  title={Deep residual learning for image recognition},
  author={He, Kaiming and Zhang, Xiangyu and Ren, Shaoqing and Sun, Jian},
  booktitle={Proceedings of the IEEE conference on computer vision and pattern recognition},
  pages={770--778},
  year={2016}
}

@inproceedings{ecapatdnn, 
   title={ECAPA-TDNN: Emphasized Channel Attention, Propagation and Aggregation in TDNN Based Speaker Verification},
   DOI={10.21437/interspeech.2020-2650},
   booktitle={Proc. Interspeech}, 
   author={Desplanques, Brecht and Thienpondt, Jenthe and Demuynck, Kris},
   year={2020},
   pages={3830–3834},
   collection={interspeech_2020} 
}

@inproceedings{pcf,
  title={PCF: ECAPA-TDNN with progressive channel fusion for speaker verification},
  author={Zhao, Zhenduo and Li, Zhuo and Wang, Wenchao and Zhang, Pengyuan},
  booktitle={Proc. ICASSP}, 
  pages={1--5},
  year={2023},
  organization={IEEE}
}

@inproceedings{nexttdnn,
  author={Heo, Hyun-Jun and Shin, Ui-Hyeop and Lee, Ran and Cheon, YoungJu and Park, Hyung-Min},
  booktitle={Proc. ICASSP}, 
  title={NeXt-TDNN: Modernizing Multi-Scale Temporal Convolution Backbone for Speaker Verification}, 
  year={2024},
  volume={},
  number={},
  pages={11186-11190},
  doi={10.1109/ICASSP48485.2024.10447037},
  ISSN={2379-190X},
}

@article{resnet-but,
  title={But system description to voxceleb speaker recognition challenge 2019},
  author={Zeinali, Hossein and Wang, Shuai and Silnova, Anna and Mat{\v{e}}jka, Pavel and Plchot, Old{\v{r}}ich},
  journal={arXiv preprint arXiv:1910.12592},
  year={2019}
}

@INPROCEEDINGS{resnext,
  author={Zhou, Tianyan and Zhao, Yong and Wu, Jian},
  booktitle={2021 IEEE Spoken Language Technology Workshop (SLT)}, 
  title={ResNeXt and Res2Net Structures for Speaker Verification}, 
  year={2021},
  volume={},
  number={},
  pages={301-307},
  doi={10.1109/SLT48900.2021.9383531}}

@inproceedings{res2net,
  title     = {{An Enhanced Res2Net with Local and Global Feature Fusion for Speaker Verification}},
  author    = {Yafeng Chen and Siqi Zheng and Hui Wang and Luyao Cheng and Qian Chen and Jiajun Qi},
  year      = {2023},
  booktitle = {Proc. Interspeech},
  pages     = {2228--2232},
  doi       = {10.21437/Interspeech.2023-1294},
  issn      = {2958-1796},
}

@inproceedings{df-resnet,
  title     = {{DF-ResNet: Boosting Speaker Verification Performance with Depth-First Design}},
  author    = {Bei Liu and Zhengyang Chen and Shuai Wang and Haoyu Wang and Bing Han and Yanmin Qian},
  year      = {2022},
  booktitle = {Proc. Interspeech},
  pages     = {296--300},
  doi       = {10.21437/Interspeech.2022-484},
  issn      = {2958-1796},
}

@ARTICLE{gemini,
  author={Liu, Tianchi and Lee, Kong Aik and Wang, Qiongqiong and Li, Haizhou},
  journal={IEEE/ACM}, 
  title={Golden Gemini is All You Need: Finding the Sweet Spots for Speaker Verification}, 
  year={2024},
  volume={32},
  number={},
  pages={2324-2337},
  doi={10.1109/TASLP.2024.3385277}}

@inproceedings{cam,
  title     = {{CAM++: A Fast and Efficient Network for Speaker Verification Using Context-Aware Masking}},
  author    = {Hui Wang and Siqi Zheng and Yafeng Chen and Luyao Cheng and Qian Chen},
  year      = {2023},
  booktitle = {Proc. Interspeech},
  pages     = {5301--5305},
  doi       = {10.21437/Interspeech.2023-1513},
  issn      = {2958-1796},
}

@inproceedings{ecapa2,
  author={Thienpondt, Jenthe and Demuynck, Kris},
  booktitle={2023 IEEE Automatic Speech Recognition and Understanding Workshop (ASRU)}, 
  title={ECAPA2: A Hybrid Neural Network Architecture and Training Strategy for Robust Speaker Embeddings}, 
  year={2023},
  volume={},
  number={},
  pages={1-8},
  doi={10.1109/ASRU57964.2023.10389750}
  }

@inproceedings{mfa-conformer,
      title={MFA-Conformer: Multi-scale Feature Aggregation Conformer for Automatic Speaker Verification}, 
      author={Yang Zhang and Zhiqiang Lv and Haibin Wu and Shanshan Zhang and Pengfei Hu and Zhiyong Wu and Hung-yi Lee and Helen Meng},
      booktitle={Proc. Interspeech},
      year={2022},
      pages={306-310},
}

@INPROCEEDINGS{transformer-speaker,
  author={Wang, Rui and Ao, Junyi and Zhou, Long and Liu, Shujie and Wei, Zhihua and Ko, Tom and Li, Qing and Zhang, Yu},
  booktitle={Proc. ICASSP}, 
  title={Multi-View Self-Attention Based Transformer for Speaker Recognition}, 
  year={2022},
  volume={},
  number={},
  pages={6732-6736},
  doi={10.1109/ICASSP43922.2022.9746639}}

@ARTICLE{cnn-transformer,
  author={Choi, Jeong-Hwan and Yang, Joon-Young and Chang, Joon-Hyuk},
  journal={IEEE/ACM}, 
  title={Efficient Lightweight Speaker Verification With Broadcasting CNN-Transformer and Knowledge Distillation Training of Self-Attention Maps}, 
  year={2024},
  volume={32},
  number={},
  pages={4580-4595},
  doi={10.1109/TASLP.2024.3463491}}

@inproceedings{redimnet,
  title     = {{Reshape Dimensions Network for Speaker Recognition}},
  author    = {Ivan Yakovlev and Rostislav Makarov and Andrei Balykin and Pavel Malov and Anton Okhotnikov and Nikita Torgashov},
  year      = {2024},
  booktitle = {{Proc. Interspeech}},
  pages     = {3235--3239},
  doi       = {10.21437/Interspeech.2024-2116},
  issn      = {2958-1796},
}

@article{redimnet2,
    title={ReDimNet2: Scaling Speaker Verification via Time-Pooled Dimension Reshaping}, 
    author={Ivan Yakovlev and Anton Okhotnikov},
    year={2026},
    journal={arXiv preprint arXiv:2603.11841},
    url={https://arxiv.org/abs/2603.11841}, 
}

@inproceedings{kim25i_interspeech,
  title     = {{Rethinking Leveraging Pre-Trained Multi-Layer Representations for Speaker Verification}},
  author    = {Jin Sob Kim and Hyun Joon Park and Wooseok Shin and Sung Won Han},
  year      = {2025},
  booktitle = {{Interspeech 2025}},
  pages     = {3713--3717},
  doi       = {10.21437/Interspeech.2025-628},
  issn      = {2958-1796},
}

@INPROCEEDINGS{w2v-bert,
  author={Li, Ze and Cheng, Ming and Li, Ming},
  booktitle={Proc. ICASSP}, 
  title={Enhancing Speaker Verification with w2v-BERT 2.0 and Knowledge Distillation Guided Structured Pruning}, 
  year={2026},
  volume={},
  number={},
  pages={16462-16466},
  doi={10.1109/ICASSP55912.2026.11461727}}

@INPROCEEDINGS{3d-speaker,
  author={Chen, Yafeng and Zheng, Siqi and Wang, Hui and Cheng, Luyao and Zhu, Tinglong and Huang, Rongjie and Deng, Chong and Chen, Qian and Zhang, Shiliang and Wang, Wen and Li, Xihao},
  booktitle={Proc. ICASSP}, 
  title={3D-Speaker-Toolkit: An Open-Source Toolkit for Multimodal Speaker Verification and Diarization}, 
  year={2025},
  volume={},
  number={},
  pages={1-5},
  doi={10.1109/ICASSP49660.2025.10888389}}

@INPROCEEDINGS{moe-ptm,
  author={Li, Yishuang and Yu, Yi and Tu, Yongfeng and Deng, Shuhao and Gan, Weihao},
  booktitle={Proc. ICASSP}, 
  title={Enhancing Speaker Verification with Layer-Wise Mixture-of-Experts on Pre-Trained Models}, 
  year={2026},
  volume={},
  number={},
  pages={19032-19036},
  doi={10.1109/ICASSP55912.2026.11464967}}

@ARTICLE{ptm-moe,
  author={Li, Zhe and Mak, Man-Wai and Pilanci, Mert and Lee, Hung-Yi and Gan, Chong-Xin and Sheng, Jiabao and Meng, Helen},
  journal={IEEE Transactions on Audio, Speech and Language Processing}, 
  title={Toward a Unified Perspective on Parameter-Efficient Fine Tuning for Speaker Verification}, 
  year={2026},
  volume={34},
  number={},
  pages={2276-2289},
  doi={10.1109/TASLPRO.2026.3682068}}

@inproceedings{sphereface2,
  author={Han, Bing and Chen, Zhengyang and Qian, Yanmin},
  booktitle={Proc. ICASSP}, 
  title={Exploring Binary Classification Loss for Speaker Verification}, 
  year={2023},
  volume={},
  number={},
  pages={1-5},
  doi={10.1109/ICASSP49357.2023.10094954}
}

@inproceedings{astp,
  title     = {{Attentive Statistics Pooling for Deep Speaker Embedding}},
  author    = {Koji Okabe and Takafumi Koshinaka and Koichi Shinoda},
  year      = {2018},
  booktitle = {{Proc. Interspeech}},
  pages     = {2252--2256},
  doi       = {10.21437/Interspeech.2018-993},
  issn      = {2958-1796},
}

@inproceedings{augment,
  title     = {{Audio augmentation for speech recognition}},
  author    = {Tom Ko and Vijayaditya Peddinti and Daniel Povey and Sanjeev Khudanpur},
  year      = {2015},
  booktitle = {{Proc. Interspeech}},
  pages     = {3586--3589},
  doi       = {10.21437/Interspeech.2015-711},
  issn      = {2958-1796},
}

@ARTICLE{musan,
    title={MUSAN: A Music, Speech, and Noise Corpus}, 
    author={David Snyder and Guoguo Chen and Daniel Povey},
    year={2015},
    journal={arXiv preprint arXiv:1510.08484},
}

@ARTICLE{rir,
  author={Szöke, Igor and Skácel, Miroslav and Mošner, Ladislav and Paliesek, Jakub and Černocký, Jan},
  journal={IEEE Journal of Selected Topics in Signal Processing}, 
  title={Building and evaluation of a real room impulse response dataset}, 
  year={2019},
  volume={13},
  number={4},
  pages={863-876},
  doi={10.1109/JSTSP.2019.2917582}
}

@INPROCEEDINGS{wespeaker,
  author={Wang, Hongji and Liang, Chengdong and Wang, Shuai and Chen, Zhengyang and Zhang, Binbin and Xiang, Xu and Deng, Yanlei and Qian, Yanmin},
  booktitle={Proc. ICASSP}, 
  title={Wespeaker: A Research and Production Oriented Speaker Embedding Learning Toolkit}, 
  year={2023},
  volume={},
  number={},
  pages={1-5},
  doi={10.1109/ICASSP49357.2023.10096626}
}

@inproceedings{lmf,
  title={The Idlab Voxsrc-20 Submission: Large Margin Fine-Tuning and Quality-Aware Score Calibration in DNN Based Speaker Verification},
  author={Thienpondt, Jenthe and Desplanques, Brecht and Demuynck, Kris},
  booktitle={Proc. ICASSP},
  pages={5814--5818},
  year={2021},
  organization={IEEE}
}

@inproceedings{asnorm,
  title     = {{Analysis of Score Normalization in Multilingual Speaker Recognition}},
  author    = {Pavel Matějka and Ondřej Novotný and Oldřich Plchot and Lukáš Burget and Mireia Diez Sánchez and Jan Černocký},
  year      = {2017},
  booktitle = {{Interspeech 2017}},
  pages     = {1567--1571},
  doi       = {10.21437/Interspeech.2017-803},
  issn      = {2958-1796},
}

@inproceedings{vox2,
  title     = {{VoxCeleb2: Deep Speaker Recognition}},
  author    = {Joon Son Chung and Arsha Nagrani and Andrew Zisserman},
  year      = {2018},
  booktitle = {{Interspeech 2018}},
  pages     = {1086--1090},
  doi       = {10.21437/Interspeech.2018-1929},
  issn      = {2958-1796},
}

@INPROCEEDINGS{cnceleb,
  author={Fan, Y. and Kang, J.W. and Li, L.T. and Li, K.C. and Chen, H.L. and Cheng, S.T. and Zhang, P.Y. and Zhou, Z.Y. and Cai, Y.Q. and Wang, D.},
  booktitle={Proc. ICASSP}, 
  title={CN-Celeb: A Challenging Chinese Speaker Recognition Dataset}, 
  year={2020},
  volume={},
  number={},
  pages={7604-7608}
  }

@inproceedings{voxblink2,
  title     = {{VoxBlink2: A 100K+ Speaker Recognition Corpus and the Open-Set Speaker-Identification Benchmark}},
  author    = {Yuke Lin and Ming Cheng and Fulin Zhang and Yingying Gao and Shilei Zhang and Ming Li},
  year      = {2024},
  booktitle = {{Interspeech 2024}},
  pages     = {4263--4267},
  issn      = {2958-1796},
}

@article{pickett2000acoustics,
  title={The Acoustics of Speech Communication: Fundamentals, Speech 
         Perception Theory, and Technology},
  author={Pickett, James M and Morris, Sherrill R},
  journal={The Journal of the Acoustical Society of America},
  volume={108},
  number={4},
  pages={1373--1374},
  year={2000},
  doi={10.1121/1.1289701}
}

@inproceedings{allosaurus,
  title={Allosaurus: A Multilingual Pretrained Universal Phone Recognizer},
  author={Li, Xinjian and Metze, Florian and Mortensen, David R and 
          Black, Alan W and Xiong, Wayne},
  booktitle={Proc. Interspeech},
  year={2020}
}

@article{moe,
  title={Outrageously Large Neural Networks: The Sparsely-Gated Mixture-of-Experts Layer},
  author={Noam Shazeer and Azalia Mirhoseini and Krzysztof Maziarz and Andy Davis and Quoc V. Le and Geoffrey E. Hinton and Jeff Dean},
  journal={ArXiv},
  year={2017},
  volume={abs/1701.06538},
  url={https://api.semanticscholar.org/CorpusID:12462234}
}

\end{document}